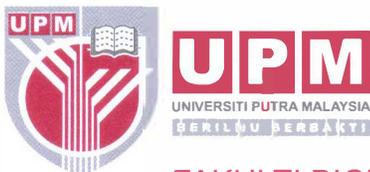

**FAKULTI BIOTEKNOLOGI DAN SAINS BIOMOLEKUL**
FACULTY OF BIOTECHNOLOGY & BIOMOLECULAR SCIENCES

4th January 2024
The editor,
Kuwait Journal of Science

Dear Sir,

COVER LETTER FOR SUBMISSION OF MANUSCRIPT

I am enclosing herewith a manuscript entitled "ISOLATION AND CHARACTERISATION OF POLYPROPYLENE MICROPLASTIC-UTILISING BACTERIUM FROM THE ANTARCTIC SOIL" for possible publication in Kuwait Journal of Science.

The paper demonstrates the ability of new Antarctic isolates in using polypropylene (PP) microplastics as their sole carbon source. This was meant to fill the gap of data scarcity in effective degradation and utilisation of microplastics in Antarctic soil remediation. As such this paper should be of interest to a broad readership including those interested in **subject areas of biodeterioration and bioavailability of microplastics in microbial ecosystems, microbial ecology and soil remediation**, which are among those targeted by the journal.

With the submission of this manuscript I would like to undertake that:

- All authors of this manuscript mutually agreed for submitting their manuscript to Kuwait Journal of Science;
- The manuscript is the original work of the authors;
- The manuscript has not been submitted earlier to Kuwait Journal of Science and;
- The contents of this manuscript will not be copyrighted, submitted, or published elsewhere, while acceptance by the Journal is under consideration;

There are no directly related manuscripts or abstracts, published or unpublished, by any authors of this paper.

- To the best of our knowledge, no conflict of interest, financial or other, exists.

Sincerely,

Dr. Nur Adeela Yasid
Department of Biochemistry,
Faculty of Biotechnology and Biomolecular Sciences,
University Putra Malaysia,
43400 UPM Serdang,
Selangor, Malaysia.
603-9649 8297
adeela@upm.edu.my

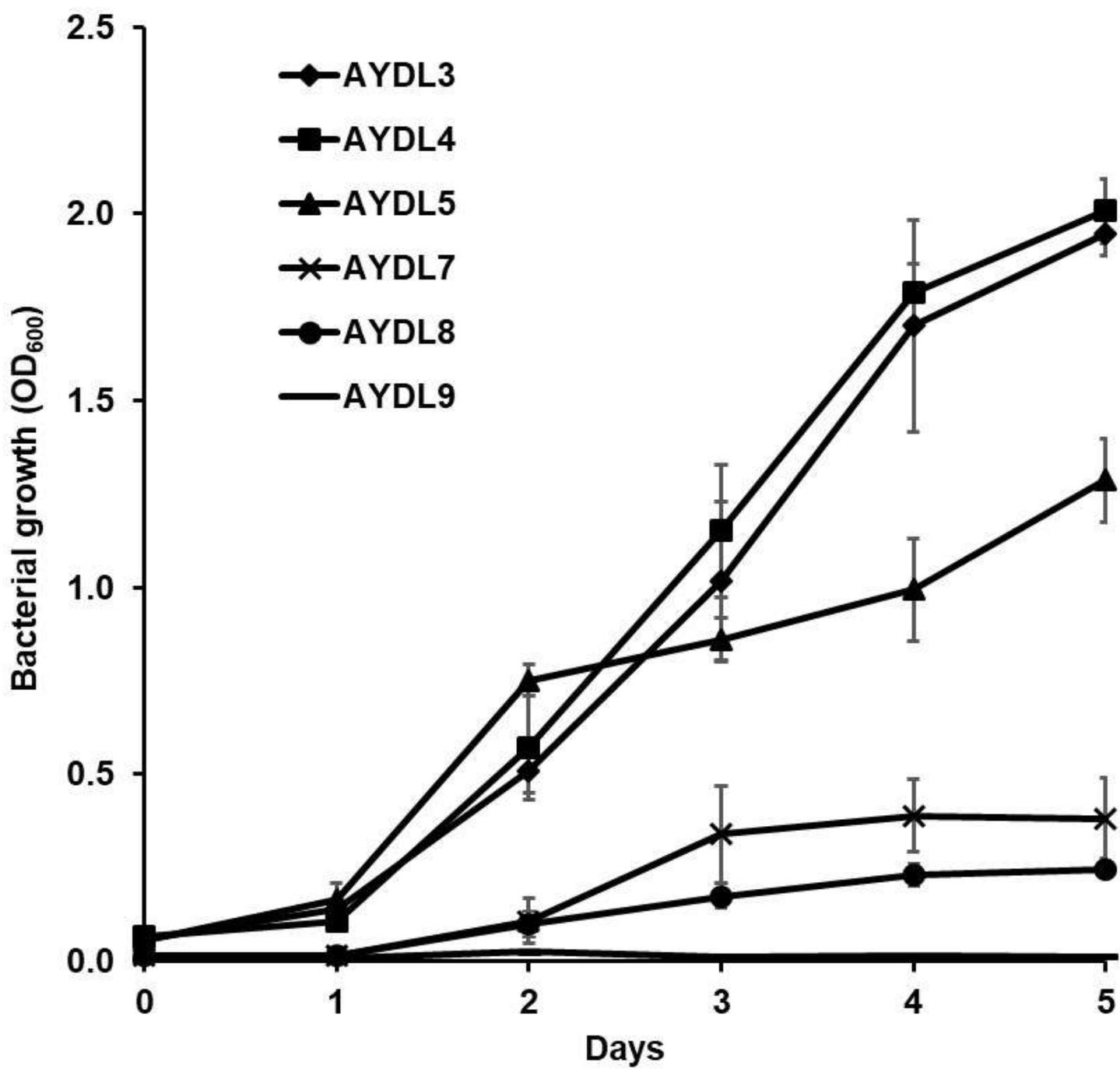

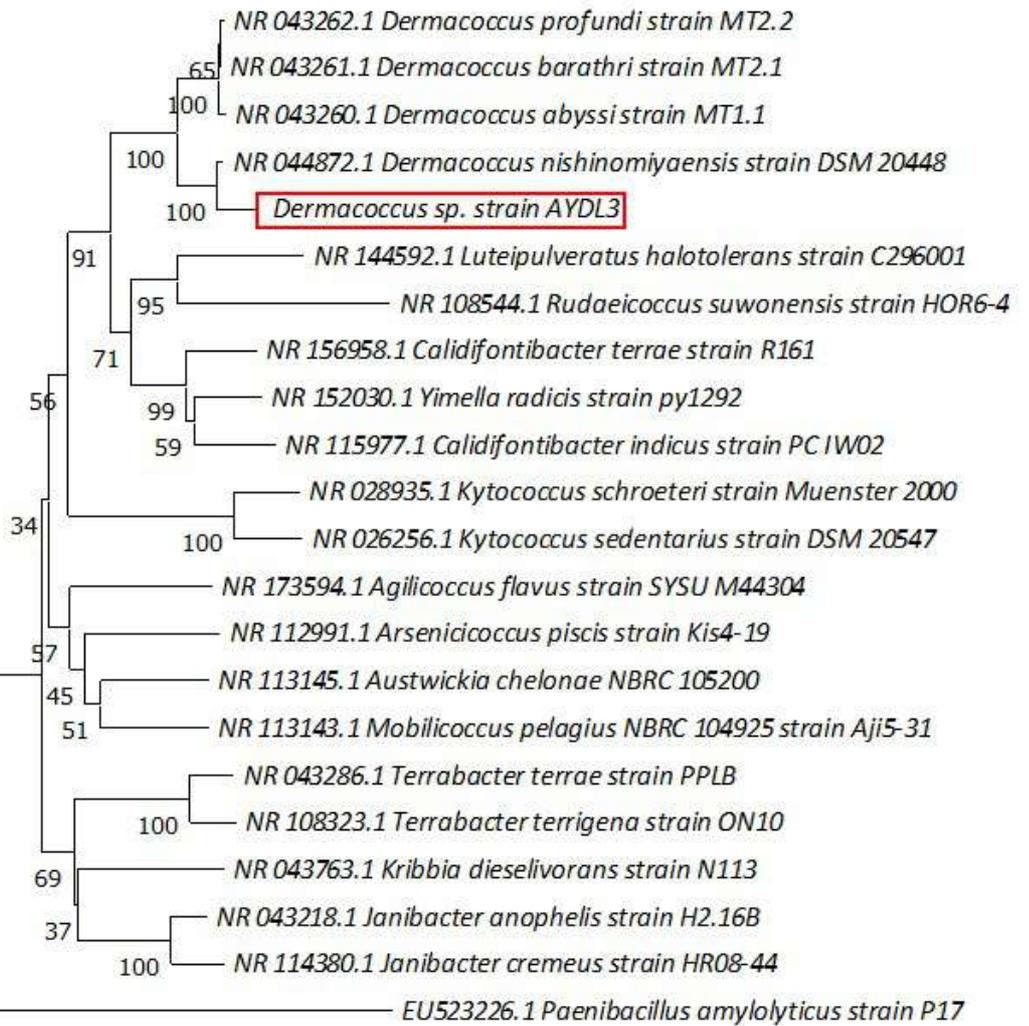

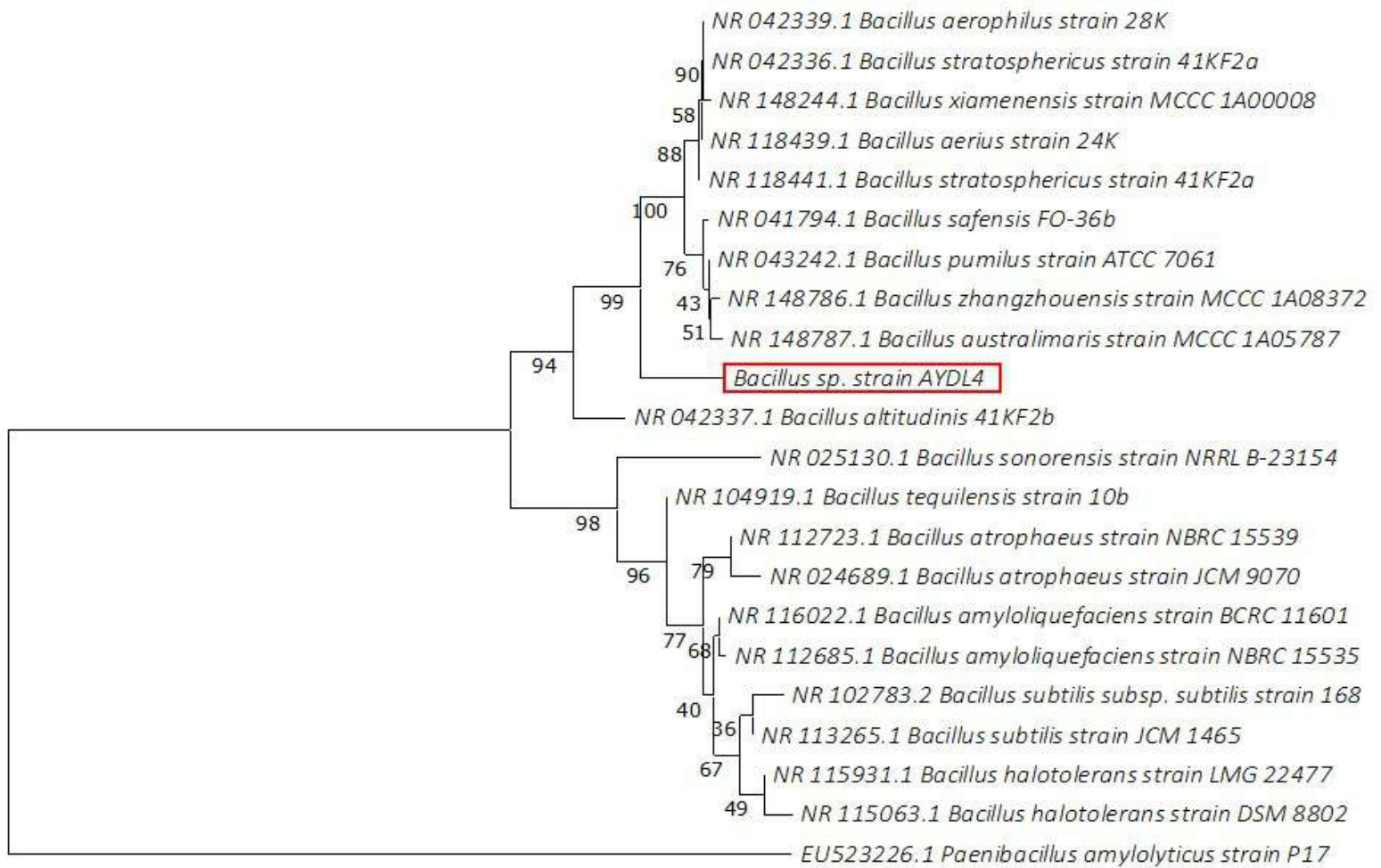

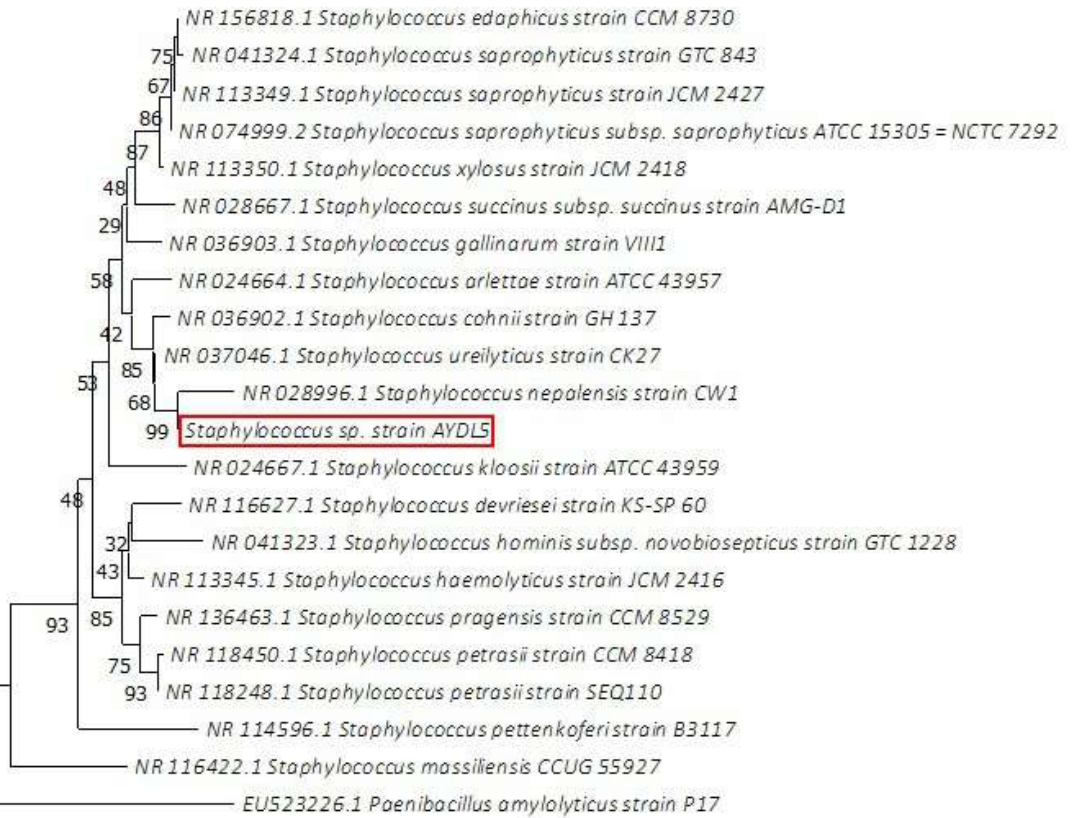

0.03

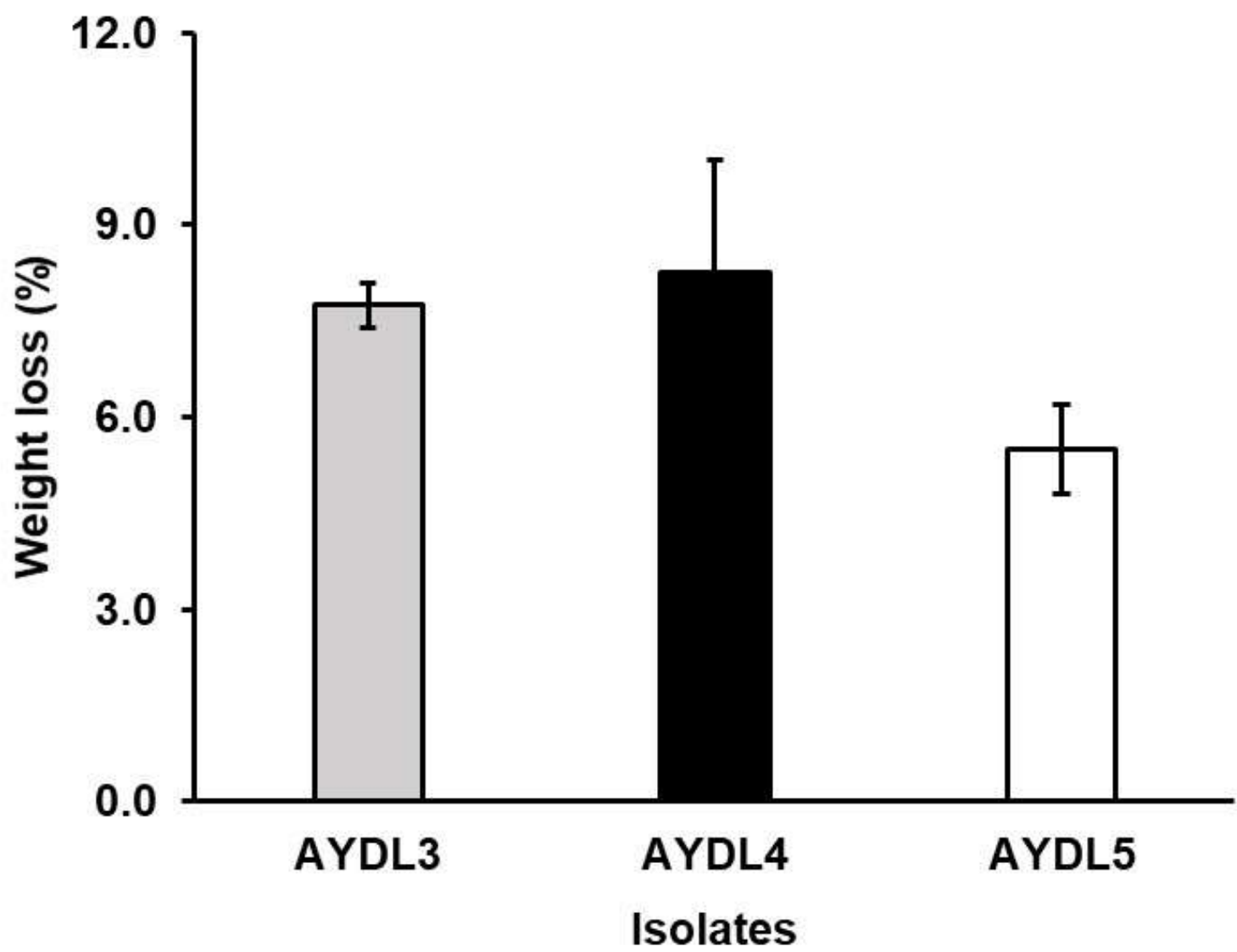

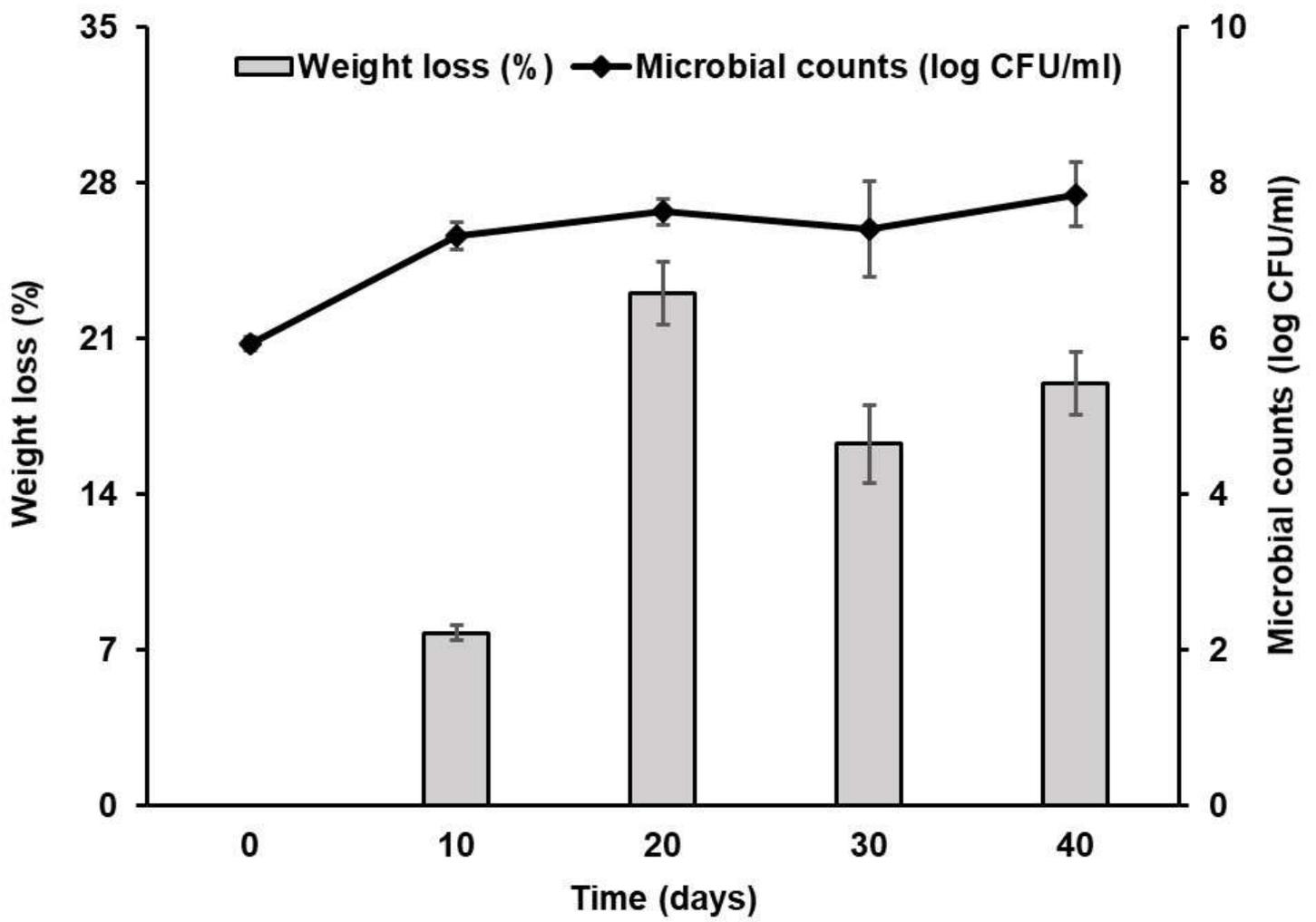

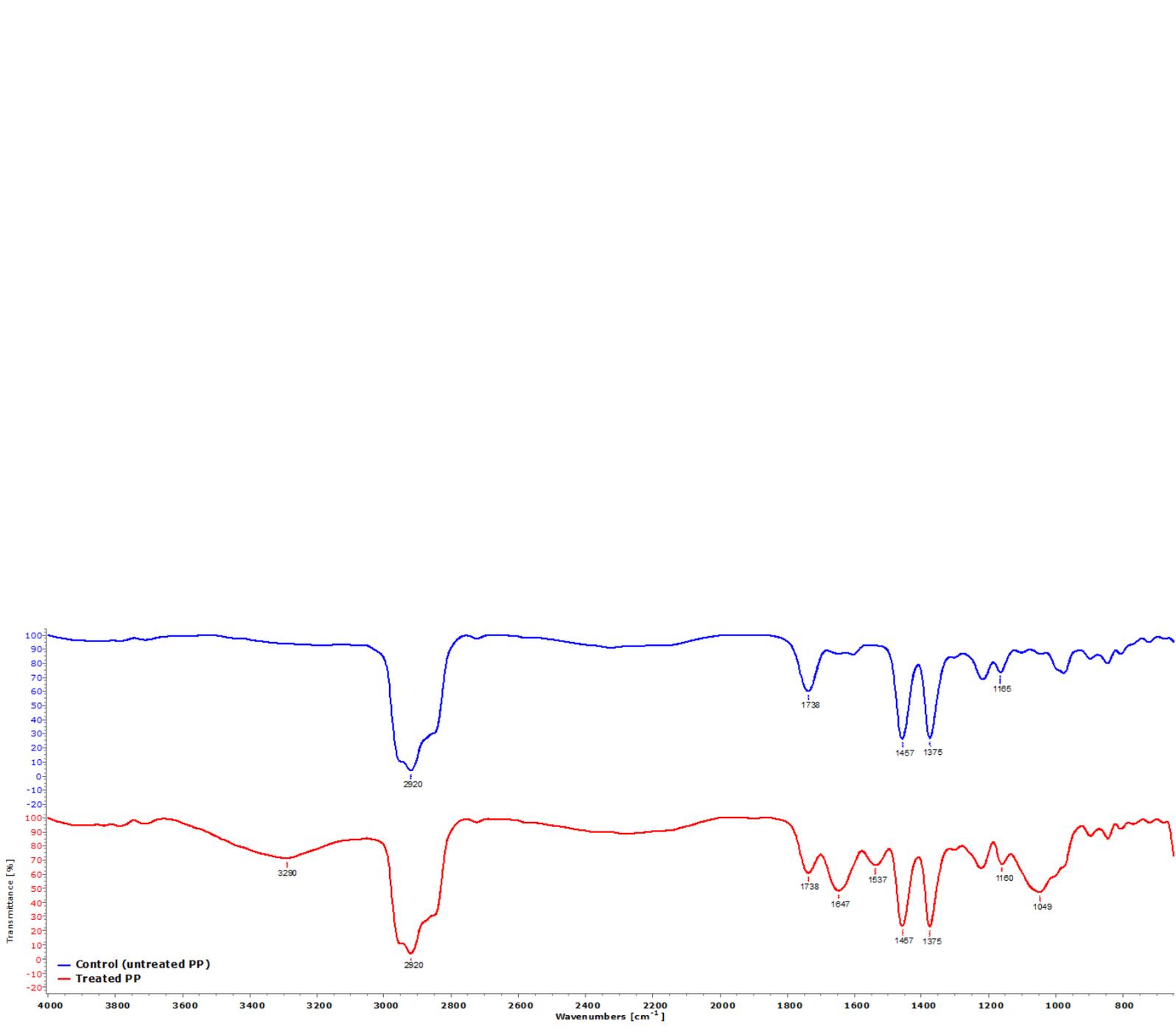

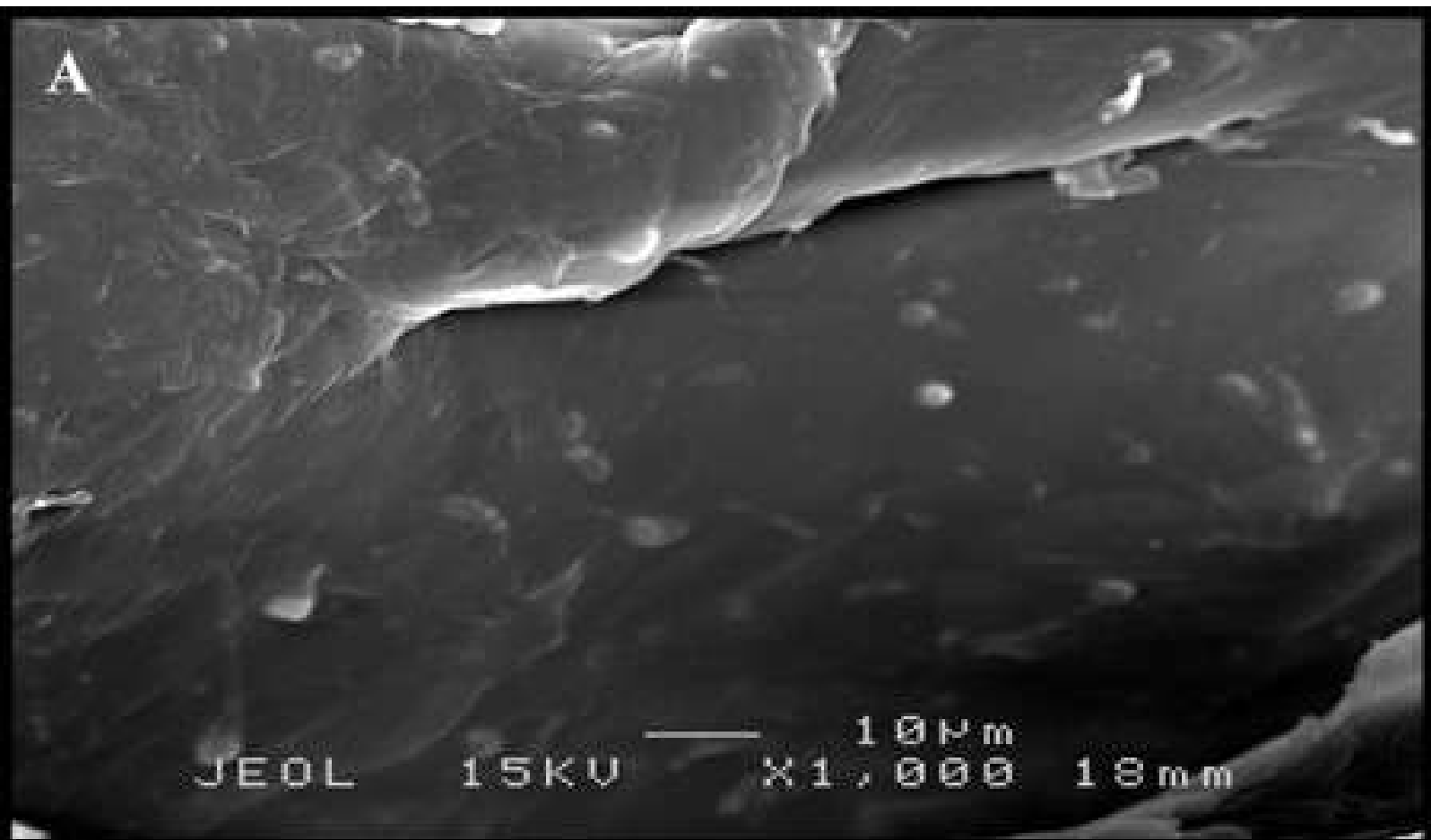
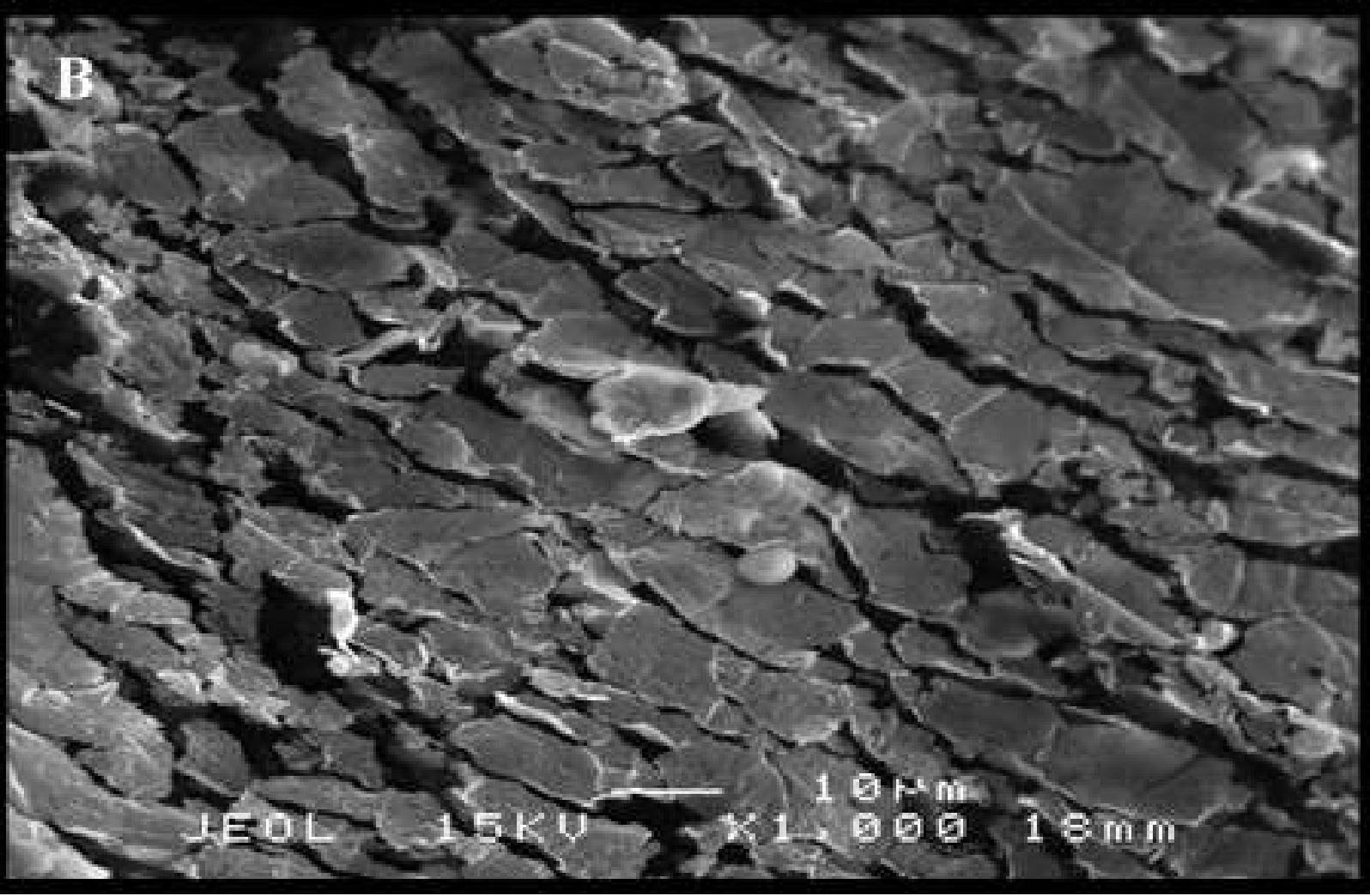

# ISOLATION AND CHARACTERISATION OF POLYPROPYLENE MICROPLASTIC-UTILISING BACTERIUM FROM THE ANTARCTIC SOIL

**Nur Ain Shuhada Ab Razak[1], Syahir Habib[1], Mohd Yunus Abd Shukor[1], Siti Aisyah Alias[2,3], Jerzy Smykla[4] and Nur Adeela Yasid[1*]**

**\*Correspondence: adeela@upm.edu.my**

[1]Department of Biochemistry, Faculty of Biotechnology and Biomolecular Sciences, 43400 UPM Serdang, Selangor, Malaysia
[2]Institute of Ocean and Earth Sciences, C308 Institute of Postgraduate Studies, University of Malaya, Kuala Lumpur, 50603, Malaysia
[3]National Antarctic Research Centre, B303 Institute of Postgraduate Studies, University of Malaya, Kuala Lumpur, 50603, Malaysia
[4]Institute of Nature Conservation, Polish Academy of Sciences, Mickiewicza 33, 31-120 Kraków, Poland

**Abstract**

Despite its remoteness from other continents, the Antarctic region cannot escape the aftermath of human activities as it is highly influenced by anthropogenic impacts that occur both in the regional and global context. Contamination by microplastics, mostly caused by the improper disposal of plastic waste, is widely recognised as a serious environmental threat due to its ubiquity. In recent years, most researchers have focused on microplastic pollution in the marine ecosystem of Antarctica, while pollution in the terrestrial environment continues to be neglected. This study was conducted to investigate the ability of Antarctic soil bacteria to use polypropylene (PP) microplastics as the sole carbon source. Bushnell Haas (BH) medium inoculated with bacteria and supplemented PP-microplastics as the sole carbon source was used in the utilisation test. In this study, the growth response of *Dermacoccus* sp. strain AYDL3 was assessed after exposure to PP-microplastics in a basal medium for 40 days. The weight reduction of the polymer was determined to further support the growth response. The highest and lowest weight loss percentages were observed on day 20 (23.0%) and day 10 (7.75%), respectively. Fourier transforms infrared (FTIR) spectroscopy and scanning electron microscopy (SEM) analyses were used to confirm the utilisation of PP-microplastics by strain AYDL3. Results indicate that the soil bacteria possess a mechanism for breaking down microplastics allowing them to utilise plastics as energy sources without any pre-treatment. This emphasises the significance of these soil bacteria to adapt and subsequently manage the plastic fragments in the soil in the future.

**Keywords:** polypropylene microplastics, utilization, weight loss, Antarctic soil, *Dermacoccus* sp.

## 1. Introduction

The small-sized plastic particles measuring <5 mm often described as 'microplastics', have posed a significant threat to the environment due to their escalating abundance resulting from large-scale manufacturing and widespread use [1,2]. Although their impact on marine ecosystems has been significantly studied, research on microplastics pollution in the terrestrial ecosystems are considerably low in contrast to the former environment. Nevertheless, recent years have shown that more attention has been shifted to the antagonistic effects of microplastics in terrestrial ecosystems, particularly soil biomes [3–8]. The disparity in research between the ecosystems can be attributed to differences in the mode of plastic accumulation, the complexity of soil composition, and challenges in extracting plastic debris from soil samples [9]. Microplastics generally enter the soil environment via improper waste disposal, domestic sewage discharge, effluents of wastewater treatment, and the conventional plastic mulching routine [4,10,11]. These synthetic wastes sequentially infiltrate the soil layers by the actions of soil biota and anthropogenic activities (harvesting, irrigation, plowing) [4,9,12]. Several types of traditional plastics have been reported to be found in the soil environment such as polypropylene (PP), polyethylene (PE), polyethylene terephthalate (PET), polyvinyl chloride (PVC), and polystyrene (PS) with each type of polymer possessing different environmental impacts [8,13].

PP is a commonly used thermoplastic made up of the propylene monomer. It is heavily utilised in a variety of industrial and consumer applications that ranges from single-use disposables to durable plastic products. While most physical and chemical characteristics of PP are similar to PE, PP has a higher melting point compared to PE rendering it a more robust, resistant, and malleable material suitable for general consumables and packaging [14]. The degradation rate and process of polymer breakdown in terrestrial ecosystems are predominantly influenced by the weathering of microplastics. While factors such as UV radiation, physical stresses (wind, abrasion, moisture), and ambient temperature minimally contribute to the deterioration of surface-level plastic waste, the restricted exposure of soil-incorporated microplastics to several of these abiotic factors, exacerbates the breakdown processes [15,16]. In light of this, microbial biodegradation emerges as a promising method to expedite the deterioration processes. Soil microorganisms, particularly bacteria exhibit an inclination to breakdown plastic polymers through the adherence of cells to plastic surfaces, followed by the subsequent colonisation of the exposed surface, and final mineralisation to carbon dioxide and water [5,16–19]. Nonetheless, as PP is hydrophobic and highly recalcitrant to degradation - owed to the long chain carbon backbone and repetitive pendant methyl groups (-$CH_3$) - efforts to boost the rate of biodeterioration have been made through physico-chemical pretreatments such as UV irradiation [20–22], thermo-oxidation [21–23], and addition of biodegradable additives and PP blends [24]. Several attempts of microbial degradation using neat or untreated PP have previously been reported with promising outcomes [25,26]. As of writing, only a single report has described the capacity of biodeterioration of PP-microplastics by bacterial strains isolated from the Antarctic. The study is distinguished for utilising hydrocarbon-degrading strains that are capable of synthesising biosurfactant - surface-active biomolecules which enhance the interaction between microbes and microplastics [27–29].

This study emphasises the ability of bacterial strain isolated from pristine Antarctic soil to grow and utilise PP-microplastics as their sole carbon source. The evaluation of bacterial strain to utilise the polymer was done by measuring the weight loss, along with assessing the structural changes using comprehensive analysis such as infrared spectroscopy and electron microscopy. This research aims to shed light on the prospect of bacteria isolated from extreme environments such as the Antarctic in utilising the recalcitrant and persistent microwaste as an energy source.

## 2. Materials and methods

### 2.1. Preparation of polypropylene (PP) microplastic

Micro-sized PP was acquired by grating the PP plastic material using a bastard-cut hand file. The grated plastic was filtered using a sieve size of 0.3 mm. The microplastic samples (<0.3 mm) were collected and sterilised with 70% ethanol for 30 minutes and dried (60°C) in an oven for 3 days prior to the experiment.

### 2.2. Bacterial isolation and preliminary screening

Soil samples were collected previously from Greenwich Island, Antarctica (62°26'51''S, 59°44'12.4''W), and maintained at -20 °C until further usage [30]. 5 g of soil sample was mixed with 50 ml of normal saline solution (0.9% w/v). The mixture was shaken for 24 h (15 ± 2 °C, 150 rpm) using an incubator shaker (LM-50RD, Yihder, Taipei, Taiwan). Aliquots (0.1 ml) of the serial dilution were spread respectively on the nutrient agar (NA) (Oxoid Ltd., Cheshire, UK) and incubated afterward for four weeks at 15 ± 2°C. All experiments were conducted in triplicates. Isolates displaying distinct colonial morphologies were isolated and further subcultured via the streak plate method. Routine cultivation was done on both NA and nutrient broth (NB) media (Oxoid Ltd., Cheshire, UK). The pure isolates were then collected from NB during their log phase and maintained in 80% v/v of sterile glycerol solution and kept at -80 °C for preservation and prevention from any contamination.

### 2.3. Secondary screening of PP utilisation by microbial isolates

For the screening process, Bushnell Haas (BH) broth (HiMedia, Mumbai, India) consisting of $MgSO_4$ (0.2 g/L), $CaCl_2$ (0.02 g/L), $K_2HPO_4$ (1.0 g/L), $KH_2PO_4$ (1.0 g/L), $NH_4NO_3$ (1.0 g/L) and $FeCl_3$ (0.05 g/L) was used throughout the experiment as the medium is devoid from any carbon sources for the bacterial growth. PP-microplastics were infused in the media to act as the only carbon source. The screening process was evaluated by inoculating 5 ml (10% v/v) of the bacterial culture ($OD_{600}$ = 0.6) into the liquid culture (45 ml BH medium with 0.2 g PP-microplastics in 250 ml conical flask). Two negative control flasks (absence of microplastic and absence of isolates) were concurrently prepared to observe any significant effects on the growth profile. The flasks were then shaken in an incubator shaker (17 ± 2 °C, 150 rpm). The weight loss of microplastic was measured and calculated on day 10. All experiments were performed in replicates.

### 2.4. Bacterial identification

The three selected isolates from the screening process undergo identification by Gram staining and 16S ribosomal RNA (rRNA) sequencing. The DNeasy Blood & Tissue Kit (Qiagen, Germany) was used in extracting genomic DNA according to the manufacturer's procedure and protocol. Extraction was done using enzymatic lysis buffer which consists of 20 mM Tris–HCl (pH 8.0), 2 mM EDTA, and 1.2% Triton X-100 with the addition of 20 mg/ml of lysozyme immediately before the experiment. Amplification of the 16S rRNA gene was performed using the universal primers: 27F (5′-AGA GTT TGA TCC TGG CTC AG-3′) and 1492R (5′-TAC GGT TAC CTT GTT ACG ACT T-3′) with a gradient thermocycler (Hercuvan, Cambridge, UK). The total mixture of PCR reaction (25 µl) consisted of 1 µl of DNA template, 1 µl of 0.4 mM forward and reverse primer each, 12.5 µl of 2X Taq Master Mix (Vivantis, Malaysia), and 9.5 µl sterile deionised water. Polymerase chain reaction (PCR) was then performed under the following conditions: initial denaturation at 94 °C for 3 min; 29 cycles of denaturation at 94 °C for 1 min, annealing at 56-59 °C for 1 min, extension at 72 °C for 2 min; and final extension at 72 °C for 10 min with incubation temperature at 4 °C. PCR products were purified using the EZ-10 Spin Column Purification Kit (BioBasic, Markham, Canada) following the manufacturer's protocol. The PCR products were separated and analysed on 1.5% w/v agarose gel electrophoresis. Finally, the purified PCR products were sequenced using a 3730xl DNA Analyzer (Applied Biosystems™, Waltham, MA, USA).

2.5. Phylogenetic analysis

The partial 16S rRNA gene sequences of each isolate were identified using BLASTn. Multiple sequence alignment was used to construct the phylogenetic tree via the neighbour-joining method with the MEGA (version 11.0) software. *Paenibaecillus amylolyticus* strain P17 was used as the outgroup for all constructed respective phylogenetic trees. The evolutionary distances were computed using the Jukes-Cantor model, while the robustness of the inferred trees was calculated and evaluated via neighbour-joining by 1000 bootstrap replicates.

2.6. Assay for PP-microplastics utilisation

BH liquid media was used during the evaluation for PP-microplastics utilisation. 5 ml (10% v/v) of the bacterial culture ($OD_{600} = 0.6$) was inoculated into a 45 ml BH medium with 0.2 g PP-microplastics in a 250 ml conical flask. A negative control flask with the absence of microplastic was prepared to observe any significant effects on the growth profile. The flasks were then shaken in an incubator shaker (17 ± 2 °C, 150 rpm). The weight loss of microplastic and microbial counts were measured and calculated every 10 days for a 40-day incubation period. All experiments were performed in replicates [20,26,30].

2.7. Determination of dry weight, reduction rate, and half-life of PP-microplastics residue.

After the incubation period of 40 days, the residual PP-microplastics polymer was obtained through filtration and subsequently washed with 70% ethanol. The sample was then left overnight in an oven at 60°C. The weight of the residual polymer was checked by using an analytical balance (Tree HRB Balance, CA, USA) with a sensitivity of 0.001 g. The percentage of weight loss of specific microplastics is calculated using Equation (1) given below.

Weight loss percentage, % = [($W_0$ - $W$) / $W_0$] ×100;

where $W_0$ is the initial weight of the microplastic (g) and $W$ is the final weight of the microplastic (g). The initial and final weight obtained was then used to calculate the rate constant of microplastic polymer reduction using the first-order kinetic model with specific intervals of 10 days [20]. The polymer reduction rate was calculated using Equation (2) as follows:

$K = -1/t$ [ln ($W/W_0$)];

where $K$ is the first-order rate constant for PP-microplastics uptake per day, $t$ refers to the time in days, $W$ refers to the final weight of PS microplastics (g), and $W_0$ refers to the initial weight of PS microplastics (g). This model and the respective formula were used to give a constant fraction of specific microplastic removal per unit of time. The half-life of a first-order reaction is independent of the original concentration of the substrate. Hence, by using the value of the microplastic removal rate constant ($K$), the half-life can be calculated using Equation (3) as follows:

$t_{1/2}$ = ln (2)/$K$

2.8. Fourier transform infrared (FTIR) analysis of PP-microplastics particles

The changes in the structure of the PP-microplastic polymers were analysed using a Spectrum 100 FTIR spectrometer (PerkinElmer, Waltham, MA, USA) in the frequency range of 4000-650 cm$^{-1}$. The types of chemical bonds (functional group) in the polymer were elucidated by using the attenuated total reflection (ATR) detector method. The spectra were constructed after conducting 16 scans with a resolution of 4.00 cm$^{-1}$. This infrared spectroscopic analysis was conducted on all samples including the uninoculated control. Prior to the analysis, the samples were collected from the BH media, retrieved via filtration, washed with 70% ethanol, and dried overnight at 60 °C in a drying oven.

2.9. Scanning electron microscope (SEM) analysis of PP-microplastics particles

The changes in the surface morphology of the control and residual microplastic were viewed after 40 days of incubation using a non-destructive scanning electron microscope (JSM 6400 SEM, JEOL Ltd., Tokyo, Japan). The samples were dispersed evenly on the carbon-coated SEM stub, sputter-coated with a gold layer at 20 mA under an argon (Ar) atmosphere at 0.3 MPa, and visualised using SEM at a magnification of 1000x.

2.10. Statistical analysis

Data analysis was conducted using the nonparametric Kruskal-Wallis test (SPSS software, version 25.0) to analyse the percentage weight loss of microplastic as a dependent variable with three different isolated bacteria as three independent groups.

## 3. Results and discussion

3.1 Isolation and primary screening of PP-microplastics utilising bacteria

Despite the extreme environmental conditions and limited nutrient accessibility, microbes are the most abundant life that colonises Antarctic ecosystems. Among the commonly studied and reported soil inhabitants in Antarctica is from the phylum Actinobacteria, due to their capability to produce novel bioactive compounds [31–34]. Previous studies have demonstrated the unique ecological characteristics of Greenwich Island where it harbours a diverse array of microorganisms, including microalgae [35], bacteria [36], and fungi [37], which possess the potential as bioremediation agents. The extreme environmental conditions of this pristine Antarctic locale may provide a distinctive niche for the evolution of microorganisms with specialised biochemical pathways, potentially enabling them to interact with and metabolise xenobiotics. Owing to that, the isolation of bacteria that are adept in utilising microplastic should be considered as the crucial step in the lengthy process to comprehend the molecular mechanisms underpinning their interactions with the said xenobiotics (PP-microplastics), contributing to a deeper understanding of the potential bioremediation applications of Antarctic microbial communities.

During the preliminary screening, six distinct bacterial isolates (AYDL3, AYDL4, AYDL5, AYDL7, AYDL8, AYDL9) were successfully isolated with similar colony morphology (circular forms with smooth margins) yet varied in their pigmentations. The pattern of bacterial growth for all isolates was observed by recording the optical density at 600 nm (OD$_{600}$) (**Figure 1**).

**Figure 1:** The growth curve of all isolates in nutrient broth (NB). All isolates were incubated for five days (15 ± 2 °C, 150 rpm). Data represent mean ± SD, $n$ = 3.

A succession of phases in the growth of a bacterial culture can be easily distinguished by the variations that occur in the growth rate. The proliferation of bacteria follows a similar pattern which is the lag phase before replication, exponential growth (active replication), and then a period of maximum population density eventually followed by the 'death phase' [38,39]. While isolates AYDL3, AYDL4, and AYDL5 were observed to be in their respective exponential phase on day 5, a slight decline in the bacterial growth of isolate AYDL7 was observed during the same period. An exponential yet very slow growth was observed for isolate AYDL8. No significant growth was observed for isolate AYDL9. Thus, the three isolates (AYDL3, AYDL4, AYDL5) that exhibited expeditious growth and higher proliferation rates were chosen to undergo further experimental processes. The selection of bacteria with better growth characteristics generally ensures more effective utilisation of available nutrients, thus facilitating a swifter progression through experimental steps. Isolated bacteria with this criterion also serve as the ideal candidate for optimisation studies as it is likely to manifest better responsive behaviour to the changes in manipulated variables (controlled laboratory settings) [40–42].

3.2 Identification of PP-microplastics utilising bacteria

The three isolates are characterised as purple-stained cocci-shaped cells indicating Gram-positive bacteria. The identification of bacteria was supported by the 16S rRNA sequence analysis. The nucleotide blast (blastn) analysis revealed that isolate AYDL3 belongs to the *Dermacoccus* genus with a high similarity percentage (>99%) against other bacterial species such as *Luteipulveratus* sp., *Rudaeicoccus* sp., *Yimella* sp., *Calidifontibacter* sp., *Kytococcus* sp., *Agilicoccus* sp., *Arsenicicoccus* sp., *Mobilicoccus* sp., *Terrabacter* sp., and *Janibacter* sp. On the other hand, isolate AYDL4 and AYDL5 showed high similarity to the *Bacillus* (>98%) and *Staphylococcus* genus (>99%), respectively. Neighbor-joining phylogenetic tree was constructed with *Paenibaecillus amylolyticus* strain P17 used as an outgroup for all respective phylogenetic trees. Strain AYDL3 was closely related to DSM 20448, a type strain of the species *Dermacoccus nishinomiyaensis,* and exhibited significant similarity to the subclade of the Mariana Trench marine *Dermacoccus* (*Dermacoccus abyssi* strain MT1.1, *Dermacoccus barathri* strain MT2.1, and *Dermacoccus profundi* strain MT2.2) (**Figure 2**). Furthermore, strain AYDL4 showed similarity with the subclade of *Bacillus aerius* strain 24 K (**Figure 3**) as both strains were located under the same clade with a bootstrap value of 99% which indicates the bolster of the node in the phylogenetic tree. On the other hand, strain AYDL5 exhibits 99% similarity with *Staphylococcus napalensis* strain CW1 (**Figure 4**). Both strains shared the same clade and were supported with significant bootstrap values indicating the closeness of their evolutionary relationship.

**Figure 2:** Phylogenetic tree of partial 16S rRNA gene sequences using neigbour-joining method for *Dermacoccus* sp. AYDL3 (□). (GenBank: OQ135226).

**Figure 3:** Phylogenetic tree of partial 16S rRNA gene sequences using neigbour-joining method for *Bacillus* sp. AYDL4 (□). (GenBank: OQ135227).

**Figure 4:** Phylogenetic tree of partial 16S rRNA gene sequences using neigbour-joining method for *Staphylococcus* sp. AYDL5 (□). (GenBank: OQ135228).

3.3 Secondary screening of PP-microplastics utilising bacteria

For the secondary screening, the selected three strains were then evaluated for their potential to utilise microplastic as their carbon sources. The carbon-lack BH medium was used with the PP-microplastics infused in the media serves as the only carbon source. Assessment of utilisation on microplastic was examined based on the weight loss percentage of PP-microplastics after 10 days of incubation (**Table 1**).

**Table 1:** Weight loss percentage of polypropylene (PP) microplastics with the presence of strain AYDL3, AYDL4, and AYDL5 after 10 days of incubation period.

| Strain(s) | Initial weight (g) | Final weight (g) | Weight loss (%) |
|---|---|---|---|
| AYDL3 | 0.2 | 0.185 ± 0.0007 | 7.75 ± 0.35 |
| AYDL4 | 0.2 | 0.184 ± 0.0040 | 8.25 ± 1.76 |
| AYDL5 | 0.2 | 0.189 ± 0.0014 | 5.50 ± 0.70 |

From **Table 1**, strain AYDL4 was observed to have the highest percentage of weight loss compared to the other two isolates. While strain AYDL5 showed a significantly lower weight loss percentage than the other isolates, the weight loss percentage between strain AYDL3 and AYDL4 did not significantly differ ($p > 0.05$) (**Figure 5**).

**Figure 5:** Percentage weight loss (%) of PP-microplastics after 10 days of incubation in Bushnell Haas (BH) media with three different isolates**.** Data represent mean ± SD, $n = 3$.

The different weight loss percentages of each isolate may imply different responses towards the infused PP-microplastics in the media. These responses could be attributed to the distinctions in metabolic rates and genetic alterations within the bacterial species [26]. Although all three isolates showed the potential in utilising PP-microplastics based on the positive weight loss, only strain AYDL3 was chosen to undergo the PP-microplastics utilisation assay. While strain AYDL5 was excluded from the utilisation assay due to the significantly lower weight loss of PP-microplastics, the omission of strain AYDL4, which was observed to have the highest weight loss percentage was based on several aspects. The weight loss percentage of AYDL4, depicted by a mean value of 8.25 with a large standard deviation of 1.76, demonstrated a considerable degree of variability among individual data points. The large standard deviation, as evidenced by the magnitude of the error bars (**Figure 2**), signifies a significant spread of values around the mean. While the mean value represents a peak in the observed range, the high variability introduces a level of uncertainty that necessitates careful consideration. In contrast, the weight loss percentage of AYDL3, with a slightly lower mean value of 7.75, exhibited a notably smaller standard deviation of 0.35. The reduced variability, as indicated by the smaller error bars, suggests a more consistent and precise dataset. While the mean value is marginally lower than that of AYDL4, the smaller standard deviation contributes to a higher level of confidence in the precision of the measurements. Due to the importance of reproducibility, precision, and the minimisation of experimental uncertainties, it was decided that the attributes of strain AYDL3 align more closely with the objectives of the subsequent stage of the study - assay for PP-microplastics utilisation. The decision to prioritise the dataset (weight loss percentage of AYDL3) with a lower mean value but a smaller standard deviation emphasises the importance of selecting data with higher precision and reduced uncertainty [43–46]. This consideration intends to enhance the reliability and robustness of the experimental outcomes, contributing to the integrity of the overall study.

3.4 Assay for PP-microplastics utilisation by *Dermacoccus* sp. strain AYDL3

The genus *Dermacoccus* was previously accredited as a xenobiotic-degrading bacteria due to its ability to degrade common soil contamination such as dichlorodiphenyltrichloroethane (DDT) and synthetic dye [47,48]. These bacteria are Gram-positive, where the cells commonly occur in irregular clusters or tetrads; and exhibit colonies that are yellowish-orange in colour. Previously identified as the genus *Micrococcus*, several species from the same genus have been isolated from extreme environments such as the deep sea Mariana Trench [49,50]. While no reports on the utilisation of PP-microplastics by *Dermacoccus* sp., the genus has been reported to be capable of degrading another type of plastic - low-density polyethylene (LDPE) [51]. Nevertheless, this study marked as the first report on the utilisation of PP-microplastics by the genus *Dermacoccus*.

To further comprehend the PP-microplastics utilisation, the bacterial growth, and tolerance were studied by incubating strain AYDL3 with the presence of the PP-microplastics for 40 days. The assay was evaluated based on the percentage weight loss (%) of PP-microplastics, and microbial counts (log CFU/ml). The highest and lowest weight loss percentages were observed on day 20 (23.0%) and day 10 (7.75%), respectively (**Figure 6**). No weight loss was observed on day 0.

**Figure 6:** Weight loss percentage of PP-microplastics and the microbial counts of *Dermacoccus* sp. strain AYDL3 in Bushnell Haas (BH) media during 40 days of incubation (15 ± 2 °C, 150 rpm). Data represent mean ± SD, $n = 3$.

The weight loss on the $10^{th}$ day incubation period may not represent whether an isolate exhibiting a higher metabolic rate but rather the isolate has a positive response towards PP-microplastics right after exposure. Bacterial cells are frequently confronted with alternative carbon sources, where they "decide" whether the available carbon sources will be preferentially consumed before metabolising the less desirable substrates (pollutants), to ensure a satisfactory metabolic return [52]. They often develop a physiological response that controls and adjusts the specific regulation and metabolic state of the cells to achieve this goal. This crucial step is essential for microorganisms to quickly adapt to the presence of new pollutants in a specific ecosystem [53]. In regard to this study, the positive metabolic response towards microplastic is believed to be the first important behavior possessed by the microbes towards the investigated polymer. This is owed to the nature of plastics which typically have extremely low bioavailability as only a very small fraction of the polymer in the early stage is exposed to potential degraders [15].

Both percentage weight loss and the microbial count showed an increase from the first ($10^{th}$ day) to the second interval ($20^{th}$ day). The steady growth of the strain AYDL3 between the intervals (0-$10^{th}$, 10-$20^{th}$) could be elucidated by the k-strategy model, where bacterial cells replicate at a slower rate while maintaining a stable environment where new progeny is produced with a greater chance of survival and persistence [28]. The substantial increment of weight loss observed on the $20^{th}$ day may imply that the strain AYDL3 has adapted to the stressed environmental condition (presence of PP-microplastics as sole carbon source), thus initiates to synthesise biofilm and excreting specific catabolic enzymes which promote the adherence of cells toward the available microplastics. Biofilm formation can be defined as a process by which bacteria attach and reproduce on the surface of the substrate. Depending on cell properties such as hydrophobicity, surface charge, or motility, biofilm development occurs to varying degrees [54]. Hydrophobic interactions appear to be crucial in the early stages of bacterial adhesion. Repulsion can occur between a negatively charged bacterium and a hydrophobic membrane. As a result, bacteria with more negative charges would be more repellent. Pang et al. [55] demonstrated that bacteria with lower cell surface charges, such as *Dermacoccus* sp., adhere more easily. Once adhesion has occurred, microorganisms with twitching or swarming motility may colonise membrane surfaces more quickly. Extracellular enzymes such as esterase, lipase, lignin peroxides, laccase, and manganese peroxides are required to increase the hydrophilicity of plastic polymers as attachment of microbes on the surface of polymer cannot happen if polymer remains its hydrophobic state [56,57]. Therefore, the insertion of new functional groups or conversion into another functional group such as carbonyl groups, can increase its hydrophilicity and allow easier microbial attachment - leading to further biodegradation in the following days [58–60].

During the collection of residual PP-microplastics from the sample flask after 30 days (3rd interval) of incubation, 'clumps' or bacterial cell aggregates were observed. It is believed that the aggregation of cells observed in the shake-flask may attributed to the decrease in the percentage weight loss of the PP-microplastics. In the scope of microplastic biodeterioration, the aggregation characteristic is often associated with the formation of biofilm – induced by bacterial cells [17,61]. Nevertheless, this aggregation characteristic is also possible to occur due to the bacteria itself trying to survive in the new environment as the condition may possess various types of environmental stress such as limited oxygen availability, or sudden temperature change projected towards the bacteria community [62]. In this study, strain AYDL3 was believed to survive by forming aggregated cells. This may indicate that strain AYDL3 employs the survival approach by the sudden change of metabolic activity, or the strain could face limited nutrients to sustain and continue growing at the higher metabolic rate - which consequently hindered the process of utilisation of PP-microplastics.

The stagnant microbial growth from the second to the last interval – concurrent with the marginal variations of percentage weight loss might signify the onset of toxic waste build-up which halted the utilisation of PP-microplastics by the remaining and active bacterial cell. It could also indicate the strain's degree of tolerance towards the PP-microplastics itself, where the strain initially demonstrated the capability to utilise the given carbon source. However, the notable shift may have occurred over time, and the bacterial cells displayed a diminishing ability to continue utilising the carbon source [63,64]. This observation suggests a time-dependent tolerance limitation, emphasising a dynamic feature of bacterial metabolism in response to the availability and utilisation of particular carbon sources.

The reduction rate, $K$ (day$^{-1}$) was determined to examine the fraction of the microplastics removed during each of the interval periods of the experiment (10-day intervals within 40 days of experimental setup). On the other hand, the half-life signifies the time needed for half of the microplastic to be reduced. The assay showed that AYDL3 recorded the highest removal rate of 0.0131 day$^{-1}$ and the shortest half-life of approximately 53 days during the second interval (20th day) with a percentage weight loss of 23% (**Table 2**). This may suggest that the best interaction between the bacteria and microplastics occurred during this particular interval. This study recorded the highest utilisation/degradation of PP-microplastics with regards to the percentage weight loss and removal rate constant; and the shortest half-life in comparison to previous reports on the utilisation of PP-microplastics by bacterial strains [20,26]. The different outcomes (higher weight loss and reduction rate, shorter half-life) are believed to be influenced by the utilisation of reduced weight of PS microplastics, and higher incubation temperature compared to the report by Auta et al. [20] and Habib et al. [26], respectively.

**Table 2:** Removal rate constant and half-life of PP-microplastics incubated with strain AYDL3.

| Days | Initial weight (g) | Final weight (g) | Weight loss (%) | Removal rate constant, $K$ (day$^{-1}$) | Half-life (days) |
|---|---|---|---|---|---|
| 0 | 0.200 | 0.200 | 0 | 0 | ∞ |
| 10 | 0.200 | 0.185 ± 0.005 | 7.75 ± 0.354 | 0.0079 | 87.7 |
| 20 | 0.200 | 0.154 ± 0.003 | 23.00 ± 1.414 | 0.0131 | 52.9 |
| 30 | 0.200 | 0.168 ± 0.004 | 16.25 ± 1.767 | 0.0058 | 119.5 |
| 40 | 0.200 | 0.161 ± 0.003 | 19.00 ± 1.414 | 0.0054 | 128.4 |

.

3.5. FTIR spectroscopic analysis of PP-microplastics

FTIR spectroscopic analysis was carried out to elucidate the utilisation of PP-microplastics by strain AYDL3 by observing any structural changes in the functional group of the polymer. Both the untreated (negative control) and treated samples (PP-microplastics incubated for 40 days with the presence of strain AYDL3) were analysed to compare any distinction in their chemical structures (**Figure 7**).

**Figure 7:** Fourier transform infrared (FTIR) spectra of control (uninoculated) PP-microplastics and treated PP-microplastics after incubation with *Dermacoccus* sp. AYDL3 for 40 days in Bushnell Haas (BH) media.

Several significant peaks from the infrared spectrum of the control sample can be observed such as C-H alkyl stretch (2920 cm$^{-1}$), C=O carbonyl bands (1738 cm$^{-1}$), and C-O phenolic bands (1165 cm$^{-1}$). Two sharp and apparent absorption peaks were also observed at 1457 and 1375 cm$^{-1}$, which corresponds to the single bonds of the C-H bend of methylene (CH$_2$) and methyl (CH$_3$) respectively. The absorption peak between 800-900 cm$^{-1}$ signifies the C-H alkyl bend of the pristine PP-microplastics. Similar infrared spectra have been reported for the pristine and untreated PP-microplastics [20,26]. While the treated PP-microplastics sample retained the previous substantial absorption peaks similar to the control samples, several new and significant absorption peaks appeared at 3290, 1647, 1537, and 1049 cm$^{-1}$. The absorption peak at 3290 cm$^{-1}$ is generally attributed to the O-H hydroxyl stretch, while the peak at 1647 cm$^{-1}$ could be associated with the presence of C=O carbonyl groups. The presence of these two new functional groups is believed to be responsible for altering the polarity of the polymer, making the PP-microplastics to be more hydrophilic – a positive indication for microbial attachment of the studied polymer. A hydrophilic surface on PP-microplastic could make it more favourable to bacterial adhesion, facilitate bacterial attachment, and the subsequent utilisation of the microplastics [61,65]. The findings of the O-H hydroxyl stretch and C=O carbonyl bands in the treated samples greatly resemble the results of the UV-treated PP-microplastics [20], and the naturally weathered PP fragments [66,67]. The emergence of the peak at 1537 cm$^{-1}$ could indicate the formation of the vinyl group (C=C). The formation of vinyl groups in polypropylene is often associated with the unsaturation of the double bond. The introduction of unsaturation can weaken the polymer structure, causing it to be more susceptible to degradation [67,68]. The new peak around 1049 cm$^{-1}$ is often associated with C-H bending vibrations. In the context of the studied polymer - polypropylene, this peak could signify the C-H bending in the polymer backbone. The changes in this region may denote alterations to the polymer structure, possibly due to bacterial degradation.

3.6. SEM analysis of PP-microplastics

The utilisation of PP-microplastics by AYDL3 was further validated by examining the morphological changes of the polymers via SEM. SEM micrograph of the control sample displayed no notable damage indicated by the smooth and intact surface without any significant surface roughness (**Figure 8A**). By comparison, numerous fissures were observed on the surface of the microplastics incubated with AYDL3 (**Figure 8B**). The rough and porous surface of the treated PP implies the utilisation of the polymer by AYDL3. Bacterial cells generally colonise and adhere to the polymer surface before the formation of biofilm and the secretion of extracellular enzymes responsible for the 'cracking' process. The subsequent formation of the pores may then initiate the metabolic utilisation of the plastic by the cells [20]. Previous studies have reported similar SEM analyses on uninoculated and inoculated PP-microplastics [20,25].

**Figure 8:** Scanning electron microscope (SEM) micrographs of (A) control (uninoculated) PP-microplastics and (B) treated PP-microplastics after incubation with *Dermacoccus* sp. AYDL3 for 40 days in Bushnell Haas (BH) media.

4. Conclusions

This study elucidates the polypropylene microplastic utilisation potential of psychrotrophic bacteria isolated from pristine Antarctic soil. Three isolates that exhibited the capability of utilising PP-microplastics – expressed by their positive response to the weight loss of PP-microplastics, were identified and designated as *Dermacoccus* sp. strain AYDL3, *Bacillus* sp. strain AYDL4, and *Staphylococcus* sp. strain AYDL5. Among the inspected strains, AYDL3 exhibited the most efficient and promising utilisation of PP-microplastics. The evaluation of bacterial strain to utilise the polymer was done by measuring the weight loss and estimating both the reduction rate and half-life of the PP-microplastics; employing the FTIR for chemical analysis, and utilising the SEM to inspect the surface morphological modifications of the investigated polymer. The results from the assessments – a discernible weight loss of PP-microplastics, the occurrence of more polar (hydroxyl and carbonyl) functional groups in the residual polymer, and apparent surface modification, demonstrated a great degree of consistency, with each method revealing concordant outcomes for a definite utilisation of PP-microplastics. This alignment of findings across different analytical approaches strengthens the reliability and robustness of the conclusions drawn from this investigation.

The commercial PP was deliberately used in the study to simulate the real-world setups of plastic pollution encountered in environmental settings. The employment of commercial PP could allow for a closer representation of the complexities present in the environment, where polymers are commonly found in various forms and compositions. By studying the interactions of the isolated *Dermacoccus* sp. strain AYDL3 with commercially available PP, this study intends to gain insights that directly resonate with practical applications and contribute to a comprehensive understanding of microbial responses to polymeric materials in environmental contexts. While the genus *Dermacoccus* has been previously reported for its capacity to degrade low-density polyethylene (LDPE), the present study discloses a new facet in the metabolic versatility of this genus – utilisation of neat PP as the sole carbon source. This finding highlights the undeveloped nature of the *Dermacoccus* genus, insinuating the vast reservoir of information that requires further exploration of the *Dermacoccus* species which could yield valuable insights into understanding and harnessing the biodeterioration capacities in the scope of polymer waste.

To improve the field of microplastic biodeterioration and biodegradation studies towards a more comprehensive system, future research initiatives could adopt a multifaceted approach. Future studies could be amended with the employment of bacterial consortia or multispecies communities to degrade microplastics instead of single bacterial strains as they effectively reflect the complex microbial dynamics encountered in natural settings – ensuring a comprehensive microplastics mineralisation process. Probing into specific metabolic pathways induced by microbes during polymer interactions in prospective research might hold great promise. This is because different genera or species could differ in their specific response towards the substrates (polymers). The involvement of numerous processes such as bacterial attachment, biofilm formation, and the subsequent enzymatic synthesis and reactions, requires advanced analytical techniques for a thorough investigation. The integration of bioinformatics, transcriptome analysis, and metabolomics as advanced analytical tools could provide a robust framework for deciphering these complexities. Lastly, optimising experimental parameters, such as temperature, incubation period, and substrate concentration could aid in enhancing the overall process of biodeterioration which is practical to real-world environmental scenarios. Collectively, these proposed augmentations might promote a more holistic exploration of microplastic-microbe interactions, creating more significant contributions to the broader field of environmental science and plastic waste management.

Acknowledgments

This work is supported by the Yayasan Penyelidikan Antartika Sultan Mizan (YPASM) Research Grant (6300303) under the Smart Partnership Initiative 2020 and the Ministry of Higher Education Research Grant (IOES-2014G).Conflicts of Interest

The authors declare no conflict of interest. The funders had no role in the design of the study; in the collection, analysis, or interpretation of data; in the writing of the manuscript; or in the decision to publish the results.

References

[1] A.I. Osman, M. Hosny, A.S. Eltaweil, S. Omar, A.M. Elgarahy, M. Farghali, P.-S. Yap, Y.-S. Wu, S. Nagandran, K. Batumalaie, S.C.B. Gopinath, O.D. John, M. Sekar, T. Saikia, P. Karunanithi, M.H.M. Hatta, K.A. Akinyede, Microplastic sources, formation, toxicity and remediation: A review, Environ. Chem. Lett. 21 (2023) 2129–2169. https://doi.org/10.1007/s10311-023-01593-3.
[2] A.A. Gazal, S.H. Gheewala, Plastics, microplastics and other polymer materials – a threat to the environment, J. Sustain. Energy Environ. 11 (2020) 113–122.
[3] S. Kublik, S. Gschwendtner, T. Magritsch, V. Radl, M.C. Rillig, M. Schloter, Microplastics in soil induce a new microbial habitat, with consequences for bulk soil microbiomes, Front. Environ. Sci. 10 (2022). https://doi.org/10.3389/fenvs.2022.989267.
[4] E.H. Lwanga, N. Beriot, F. Corradini, V. Silva, X. Yang, J. Baartman, M. Rezaei, L. van Schaik, M. Riksen, V. Geissen, Review of microplastic sources, transport pathways and correlations with other soil stressors: A journey from agricultural sites into the environment, Chem. Biol. Technol. Agric. 9 (2022) 20. https://doi.org/10.1186/s40538-021-00278-9.
[5] K.C. Omidoyin, E.H. Jho, Effect of microplastics on soil microbial community and microbial degradation of microplastics in soil: A review, Environ. Eng. Res. 28 (2023). https://doi.org/10.4491/eer.2022.716.
[6] C. Wang, J. Tang, H. Yu, Y. Wang, H. Li, S. Xu, G. Li, Q. Zhou, Microplastic pollution in the soil environment: characteristics, influencing factors, and risks, Sustainability. 14 (2022) 13405. https://doi.org/10.3390/su142013405.

Figure Captions

**Figure 1:** The growth curve of all isolates in nutrient broth (NB). All isolates were incubated for five days (15 ± 2 °C, 150 rpm). Data represent mean ± SD, *n* = 3.

**Figure 2:** Phylogenetic tree of partial 16S rRNA gene sequences using neigbour-joining method for *Dermacoccus* sp. AYDL3 (□). (GenBank: OQ135226).

**Figure 3:** Phylogenetic tree of partial 16S rRNA gene sequences using neigbour-joining method for *Bacillus* sp. AYDL4 (□). (GenBank: OQ135227).

**Figure 4:** Phylogenetic tree of partial 16S rRNA gene sequences using neigbour-joining method for *Staphylococcus* sp. AYDL5 (□). (GenBank: OQ135228).

**Figure 5:** Percentage weight loss (%) of PP-microplastics after 10 days of incubation in Bushnell Haas (BH) media with three different isolates. Data represent mean ± SD, *n* = 3.

**Figure 6:** Weight loss percentage of PP-microplastics and the microbial counts of *Dermacoccus* sp. strain AYDL3 in Bushnell Haas (BH) media during 40 days of incubation (15 ± 2 °C, 150 rpm). Data represent mean ± SD, *n* = 3.

**Figure 7:** Fourier transform infrared (FTIR) spectra of control (uninoculated) PP-microplastics and treated PP-microplastics after incubation with *Dermacoccus* sp. AYDL3 for 40 days in Bushnell Haas (BH) media.

**Figure 8:** Scanning electron microscope (SEM) micrographs of (A) control (uninoculated) PP-microplastics and (B) treated PP-microplastics after incubation with *Dermacoccus* sp. AYDL3 for 40 days in Bushnell Haas (BH) media.